Arxiv 2506.00063 , Fluct. Noise Lett. in press (2025)

# "Quantum supremacy" challenged. Instantaneous noise-based logic with benchmark demonstrations

Nasir Kenarangui [1], Walter C. Daugherity [2], Arthur Powalka [1] and Laszlo B. Kish [1,3,†]

[1] *Department of Electrical and Computer Engineering, Texas A&M University, College Station, TX 77843-3128, USA*

[2] *Department of Computer Science & Engineering, Texas A&M University, College Station, TX 77843-3112, USA*

[3] *Department of Biomatics and Artificial Intelligence, Obuda University, Budapest, Becsi ut 96/B, Budapest, H-1034, Hungary*

INVITED TALK

Instantaneous Noise-Based Logic (INBL) represents a computational paradigm that offers a deterministic alternative to quantum computing, potentially challenging the notion of quantum supremacy without relying on quantum hardware. INBL encodes logical information in orthogonal stochastic processes ("noise-bits") and exploits their superpositions and nonlinear interactions to achieve an exponentially large computational space of dimension $2^M$, where $M$ corresponds to the number of noise-bits analogous to qubits in quantum computing. This approach enables an exponential increase in computational throughput, with a computational speedup scaling on the order of $O(2^M)$, while maintaining hardware complexity comparable to quantum systems. Unlike quantum computers, INBL operates without decoherence, error correction, or probabilistic measurement, yielding deterministic outputs with low error probability. Demonstrated applications include exponential speed-gain compared to classical computers, such as INBL phonebook searches (for number or name lookup) and the implementation of the Deutsch-Jozsa algorithm, illustrating INBL's capability to perform special-purpose computations with quantum-like exponential speedup using classical-physical noise-based hardware. We present an experimental comparison between the execution speeds of a Classical Turing machine algorithm - which changes the values of odd numbers in an exponentially large set to their next lower even numbers - and its INBL counterpart. Another experimental demonstration of the exponential speedup in finding and removing a given number from an exponentially large, unsorted set of integers.

## 1. Introduction

The concept of "quantum supremacy" describes the theoretical capability of quantum computers to outperform conventional classical computers (those based on the Turing machine paradigm) for specific, computationally intensive problems. Quantum supremacy is achieved when a quantum device can solve a problem markedly faster (often exponentially so) than the best-known classical algorithms, such that the computational resources (including hardware and processing time) required by classical systems grow exponentially relative to those needed by quantum systems for the same task.

† Corresponding Author

Although the realization of universal, gate-based quantum computers, promising a computational speedup by a factor of $2^M$ at a given level of hardware complexity, remains an ongoing pursuit, where $M$ is the number of qubits utilized, recent developments have invigorated the competition between quantum and classical approaches. Advances in classical algorithms have significantly reduced the performance gap, and in some instances, have even called into question previous claims of quantum supremacy [1–6].

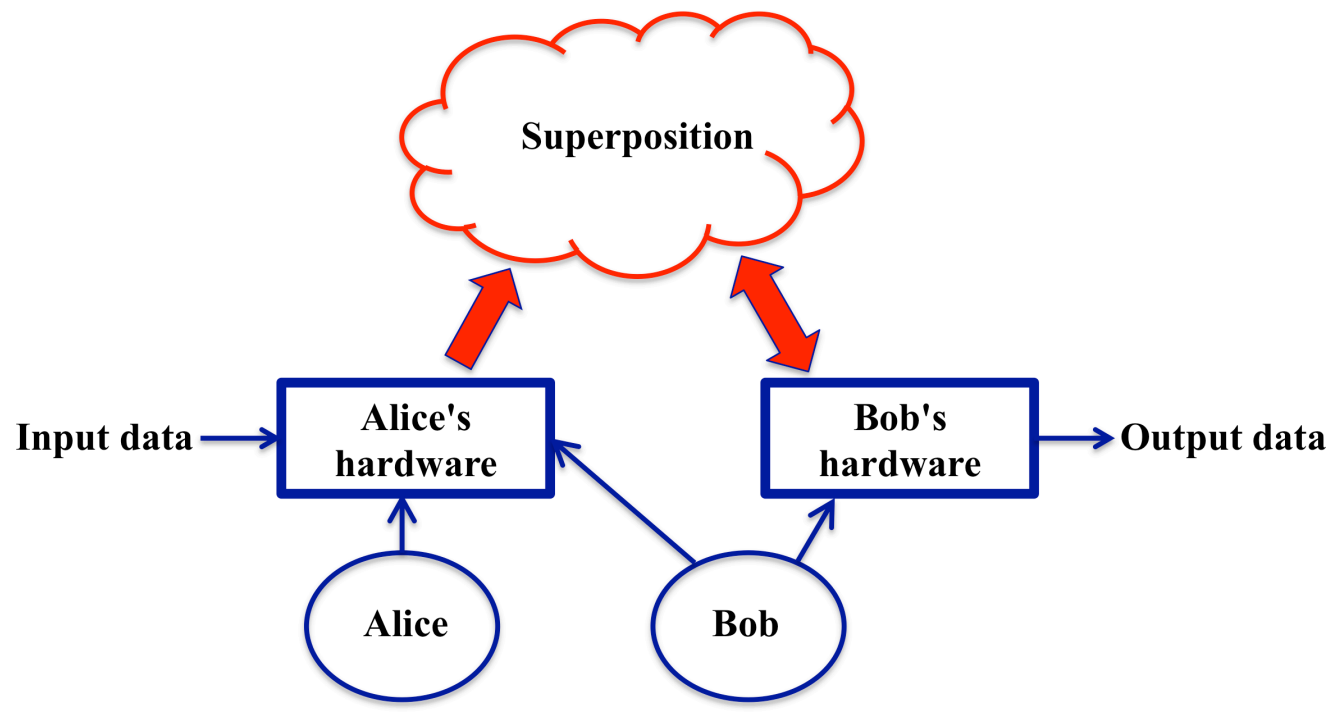

Fig. 1. Quantum computer scheme [1].

Figure 1 illustrates the architecture and operational workflow of a generic quantum computer, structured around two principal roles: the problem implementer (Alice) and the problem solver (Bob). In this paradigm, Alice is responsible for encoding the computational problem onto the quantum hardware. For certain tasks, such as populating an arbitrary database of size $K=2^M$, the resources required by Alice may scale exponentially with the parameter $M$, reflecting the exponential growth in the number of quantum states that must be initialized or manipulated.

To encode the problem, Alice prepares the quantum system in a superposition state that represents all possible inputs or configurations relevant to the problem. This preparation involves setting the initial and boundary conditions of the system's wavefunction, thereby defining the data landscape on which the quantum computation will operate. The hardware used by Alice must be capable of generating and maintaining these superpositions, leveraging the principles of quantum mechanics such as superposition and entanglement.

Subsequently, Bob executes the quantum algorithm designed to solve the encoded problem. This involves applying a sequence of quantum operations (gates) to the superposition state, possibly utilizing both the hardware configured by Alice and additional resources of her own. Upon completion of the computation, Bob performs measurements on the quantum system to extract the final results.

A key feature of an effective quantum processor is that, while Alice's task of problem encoding may require resources that scale exponentially with $M$, Bob's computational workload - i.e., the number of quantum operations and measurements needed to obtain the solution-scales only polynomially with $M$ for certain classes of problems. This polynomial scaling underlies the quantum computational advantage, enabling quantum processors to efficiently solve problems that are intractable on classical computers due to their exponential complexity.

### *1.1 Noise-based logic (NBL)*

In noise-based logic (NBL) [1,7-21], logic information is encoded using independent, zero-mean stochastic processes - referred to as noises - which act as orthogonal reference signals. For a system with $M$ noise-bits, $2M$ independent noise sources are required, as each logical bit value-High (H) and Low (L)-is represented by its own independent noise process. These reference noises are generated by the Reference Noise System [1] and distributed throughout the processor to serve as the basis for logic operations. Figure 2 typically illustrates the generic architecture of an NBL processor, where both the noise sources and the logic gates can take various forms, including correlators [7], filters [7], switches [1,7], and algebraic or set-theoretic operations [1,7-21].

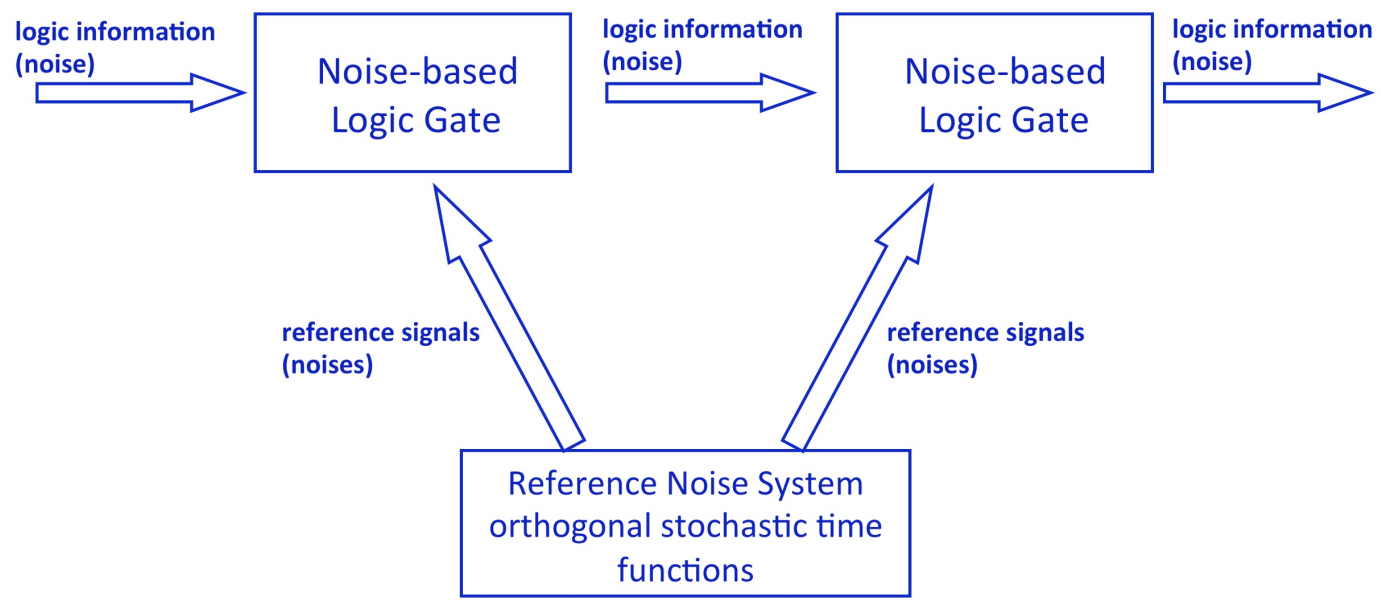

Fig. 2. Generic noise-based logic hardware scheme (e.g. [1]). Logic operations can be executed by the gates or by operations on the reference signals. The Reference Noise System is based on a truly random number generator. The interesting concrete schemes and algorithms are those can be implemented by a binary classical computer (Turing machine) with a bit resolution that is polynomial in $M$.

### *1.2 Instantaneous Noise-Based Logic (INBL)*

Instantaneous noise-based logic (INBL) [9-21] is a subclass of NBL in which logic operations are performed using dichotomous (real or complex) random telegraph waves (RTWs) or spikes, a type of stochastic process that alternates between two values (e.g., +1 and -1) with equal probability at each time step. Unlike classical NBL schemes, INBL does not require time-averaging; instead, logic results are determined instantaneously based on the current values of the noise processes. The gates in INBL systems are typically constructed using only algebraic operators and switches, omitting correlators and filters, which enables much faster computation compared to time-averaged NBL.

Figure 3 generally depicts a typical INBL architecture and Table 1 summarizes some key differences between INBL, Quantum Computing, or Classical (Turing) Computing.

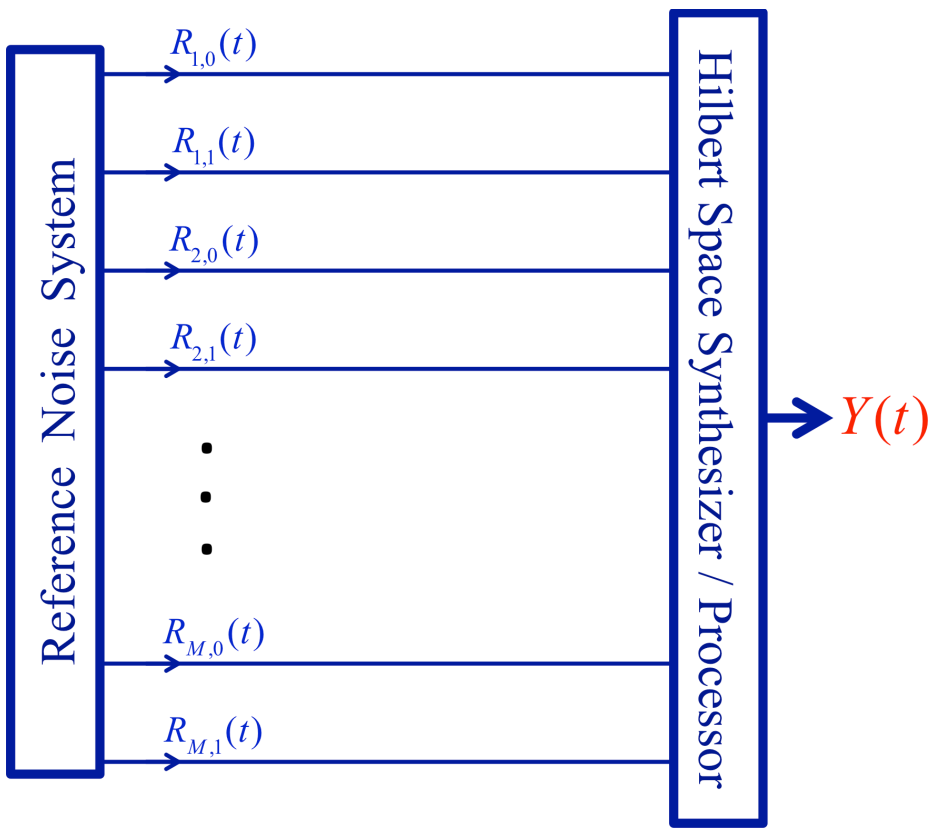

Fig. 3. Circuit illustration of the logic structure of the generic superposition synthesizer for instantaneous NBL (INBL) (e.g. [1]). Operations can take place on the exponential superposition $Y(t)$ and the references noises $R_{i,j}(t)$.

Table 1. Comparison table between INBL, Gate based quantum computers, and Classical computers with deterministic algorithm

| Feature | INBL | Quantum Computing | Classical Computing |
|---|---|---|---|
| State Space Dimension | $2^M$ (M = noise-bits) | $2^M$ (M = qubits) | $M$ |
| Core Logic Carrier | Orthogonal stochastic processes | Quantum states (superposition) | Deterministic bits |
| Output Type | Deterministic | Probabilistic (measurement) | Deterministic |
| Decoherence/Error Correction | Not required | Required | Not required |
| Hardware Complexity | Polynomial in M (classical) | Polynomial in M (quantum) | Polynomial in M |
| Speedup (Special Problems) | Exponential ($O(2^M)$) | Exponential ($O(2^M)$) | Linear or polynomial |

In INBL, the $2M$ noise sources are input into a Hilbert Space Synthesizer (a hardware or algorithm that generates the superposition of product strings, see in section 2), which generates the appropriate superpositions of noise vectors for further logic operations. This process enables the creation of a $K=2^M$ dimensional superposition - referred to as the Universe - through an operation analogous to the Hadamard transformation (applying

Hadamard gates) in quantum computing, see below. This operation, sometimes called the Achilles operation, produces a superposition of all possible integer values represented by the system, similarly to the quantum superposition of computational basis states. As a result, INBL can exploit high-dimensional logic spaces and superpositions, allowing for massively parallel information processing reminiscent of quantum computation, but using classical stochastic processes.

Figure 3 shows Polynomial circuit illustration of the logic structure of the generic superposition synthesizer of Instantaneous Noise Based Logic. The output represents a superposition of $O(2^M)$ orthogonal product strings of random telegraph waves.

## 2. INBL Engine and Implementation of bit-specific "NOT" Operations

To demonstrate the capabilities of INBL algorithms we built an INBL engine. Reference noises are generated for each bit based on an asymmetric RTW scheme [1]. The RTW is generated by random square waves with values that are positive or negative with a probability of 0.5. For $M$ noise bits, each bit is associated with 2 uncorrelated reference noises: the logical low $L_i \in \{-0.5, +0.5\}$ and the logical high $H_i \in \{-1, +1\}$, where $i = 0,1,\dots,M-1$.

For the RTWs, a specific duration was chosen to ensure, with sufficiently high probability, that all reference noises were unique. Considering that the theoretical error probability of an idealistic gate in modern computers is $10^{-25}$ [20]. RTWs of $T$ clock cycles have probability $0.5^T$ of being identical. This means we require a minimum $T = 83$ clock cycles in order to reach our target error probability $0.5^T = 10^{-25}$ [20].

Let $G_i^k(t) \in \{L_i(t), H_i(t)\}$ represent bit $i$ of the $M$ bit long binary string $\{G_1^k(t), G_2^k(t), \dots, G_M^k(t)\}$ associated with the number $k$, where $k = 0,1,\dots,2^M - 1$.

Each of the $2^M$ hyperspace vectors in the INBL system corresponds to a binary number with $M$ bit resolution. The bitstring product of reference noises for any number $k$ generates the hyperspace vectors $S_k(t)$:

$$S_k(t) = \prod_{i=0}^{M-1} G_j^k(t) \quad (1)$$

To generate the superposition of all possible integer numbers with a polynomial complexity (Achilles operation, e.g. [1]), the Universe, the H and L bit values of each bit references must be added and all these sums must be multiplied, very similarly to quantum computers, where the multiplication of such sums in INBL is replaced by Hadamard operations in quantum computers. Utilizing large superpositions of hyperspace vectors facilitates the transmission of large numerical sets through a single wire, greatly decreases memory requirements, and permits operations to be executed on these superpositions. (e.g. [1]). One such INBL tool is a very simple operation that we call "NOT" for convenience reasons:

$$\mathrm{NOT}_i = H_i(t) L_i(t) \quad (2)$$

It was introduced earlier for symmetric reference systems, and it can be applied to individual hyperspace vectors or any superposition. The $\mathrm{NOT}_i$ operation acting on the $i$-th noise-bit in INBL shown by Equation 2 is a powerful bit manipulation operation, simultaneously flipping the bit $i$ on an entire exponential superposition $U(t)$.

$$U_{i^*}(t) = \mathrm{NOT_i}\, U(t) \tag{3}$$

Where $U_{i^*}(t)$ denotes the resulting superposition, in which the noise bit $i$ of all constituent hyperspace vectors is complemented, representing a completely different set of numbers.

It is important to note that, due to the asymmetric nature of the chosen reference system, the NOT operation described above functions exactly as intended only when inverting High bit values. To invert Low bit values, the result must be multiplied by 4; this is why this operator is not a real NOT operator as with asymmetric reference systems. However, for our specific task of transforming odd numbers into even numbers (Section 4), this particular NOT operator is ideal. Moreover, for "real" NOT operations on bit values in superposition, alternative reference systems may be employed (such as complex amplitudes), or the NOT operation can also be implemented by *swapping the reference H and L wires of the respective noise-bit* [10].

## 4. Practical Algorithm Benchmarking: Classical Versus INBL

In classical computing, operations such as bitwise AND, OR, and XOR must act on logic values independently. Classical algorithms can be characterized as having step-by-step structures with element-by-element operations with logical order. This is in contrast with INBL, which allows for superpositions of an arbitrary number of logic values and simultaneous operation on these superpositions.

Several published studies have demonstrated the practical applications of INBL algorithms (e.g. [1,12,19,20]. Notably, INBL has been shown, at least theoretically, to solve certain exponentially large problems in polynomial time. Examples include the Deutsch–Jozsa problem (deciding if a bit set is uniform or symmetric); selection and removal of numbers from exponentially large, non-indexed sets (“hats”); string search within exponentially large sets; and searching for a number or name in an exponentially large phonebook.

In this paper, we implement both the Classical and INBL algorithms on the same computational platform and provide a practical demonstration for INBL in solving two representative problems. The computer on which both the INBL and Classical algorithm were run was an HP ENVY laptop with 32GB RAM.

### *4.1 Given a heap (an unsorted, unindexed collection) of odd numbers, convert each to the even number one less than itself*

Note: after carrying out this simulation, one of us (WD) found that the problem can also be solved by a special-purpose classical algorithm, thus there is no actual supremacy issue here. However, we decided to keep this material in the paper because it is one of the simplest illustration of implementing a problem using an INBL scheme.

In this problem, we have a set $O = \{k \in \mathbb{Z} | k \ is \ odd\}$ of random odd numbers. The task is to transform this set to another set $E = \{k-1 \in \mathbb{Z} | k-1 \ is \ even\}$ that will contain the even numbers. The value of each number in set $E$ is 1 less than its counterparts in set $O$. The flow chart in Figure 4 illustrates the implementation and the verification procedure.

The odd numbers are generated randomly with range of $[1, 2^M - 1]$, where $M = 32$ is the noise bit resolution used in the INBL engine. We choose the size of the set to be exponentially large $2^N$ where $N = 0,1,2, \dots$ .

Using Equation 1, each number $k$ in set $O$ is transformed into a hyperspace vector:

$$S_k(t) = \prod_{i=0}^{M-1} G_j^k(t) \qquad (4)$$

In the INBL algorithm, the set $O$ can then be represented by the superposition of these hyperspace vectors shown by:

$$U_{odd}(t) = \sum_{k=0}^{2^N-1} S_k(t) \qquad (5)$$

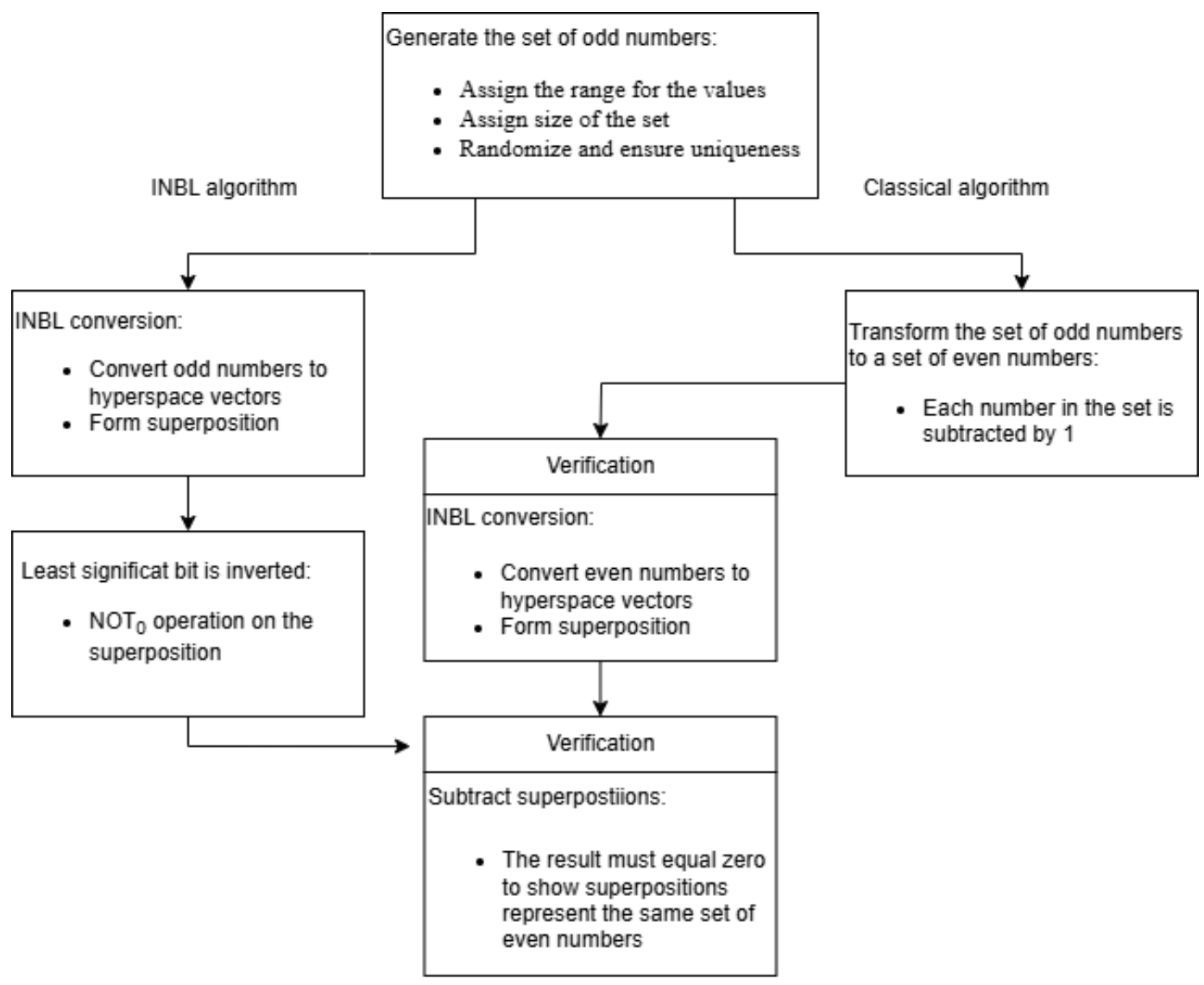

Fig. 4. Overview of the INBL and the Classical algorithms for transforming odd numbers to even, and the procedure used for verification.

The NOT operation targeting the least significant bit is defined as follows:

$$\text{NOT}_0(t) = H_0(t)L_0(t) \tag{6}$$

Finally, $\text{NOT}_0(\text{t})$ operation is executed on the superposition $U_{odd}(t)$ to obtain the superposition $U_{even}(t)$, which represents the set of even numbers $E$.

$$U_{even}(t) = \text{NOT}_0(t)\, U_{odd}(t) \tag{7}$$

For verification, $U_{even}(t)$ can then be compared with the output of the Classical algorithm for odd to even transformation. The set $E$ is converted into an INBL superposition $U^*_{even}(t)$. The comparison is done by a subtraction operation:

$$U_{even}(t) - U^*_{even}(t) = 0 \tag{8}$$

We took advantage of the Google Benchmark library for C++ developers to measure the CPU times of both the Classical and INBL implementations. This is a powerful tool for accurate measurement and analysis of code performance. Google Benchmark library achieves this by "automatic iteration" to determine how many iterations are needed to get stable and accurate timing results. The code runs repeatedly adjusting the number of runs to minimize noise and variance. For a fair and meaningful comparison, we also ensured that the measurements were carried out on the same machine and under consistent conditions, such as background processes, and operating modes in the computer. Other important considerations were programming decisions. We measured only the relevant algorithm, not the memory allocation and deallocation which were relegated outside of the function being benchmarked.

In Equation 7 the product of the $\text{NOT}_0(\text{t})$ waveform and the superposition waveform $U_{odd}(t)$ would require $T$ multiplications, which is the number of clock cycles selected for our RTWs as discussed in section 3.

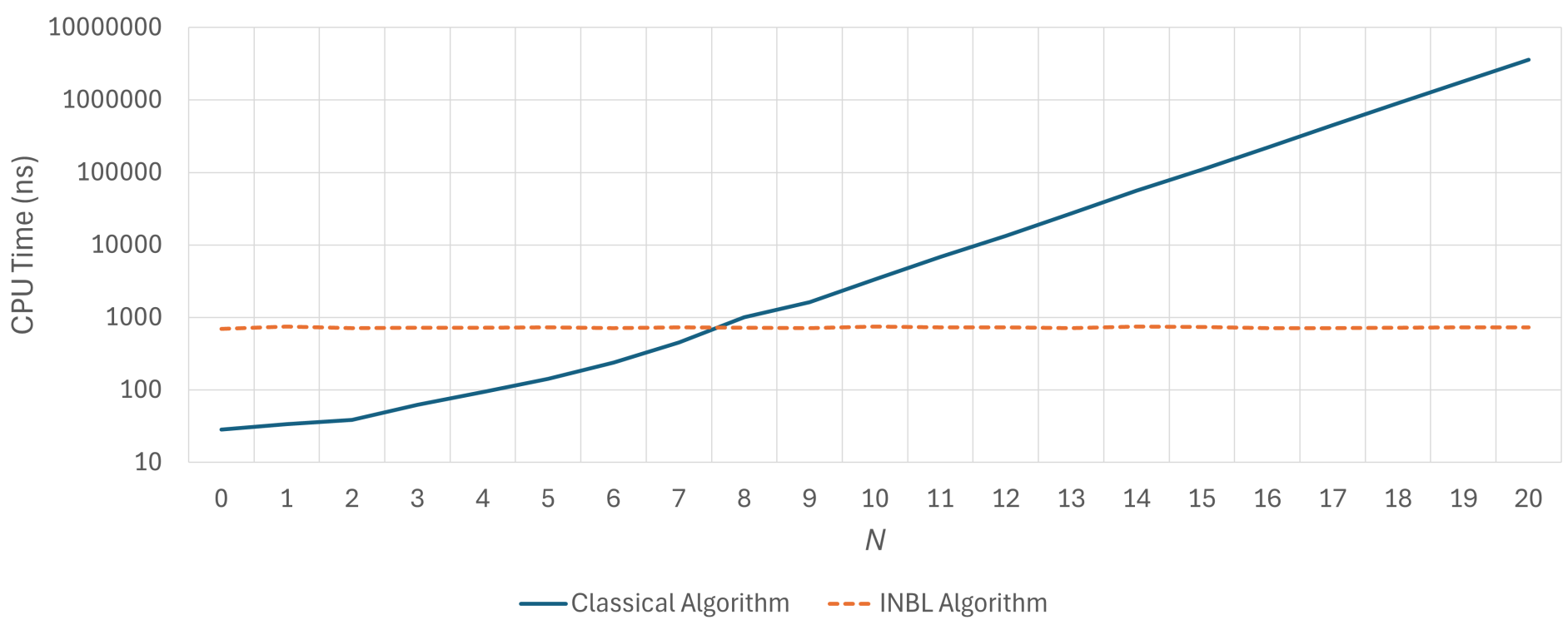

Fig. 5: The time requirement of the INBL and Classical algorithms for changing $2^N$ odd numbers to even ones. Both the hardware and time complexity of the Classical algorithm were growing exponentially with O($2^N$), while the hardware complexity of INBL scaled only polynomially, O($M$), where the number $M$ of noise-bit resolution was fixed at 32. The time complexity of INBL was constant, O(1).

For these simulations we emulated $2^{N\text{-}1}$ ($N$<$M$) odd numbers in the actual Universe and selected a fixed constant $T = 100$, resulting in time complexity $O(T) = O(1)$. That means, the Classical algorithm required storing $2^{M-1}$ odd numbers, while the INBL operated by only O(100*2*2*$M$) odd numbers, where the double-2 originates from the $2M$ noise references and from the fact that the asymmetric reference RTWs are 2-bit numbers (though in commercial computers usually 32 or 64 is the minimum size of memory allocated for a "word"). The number of operations in the Classical algorithm on the other hand is linearly dependent on $2^N$, as the algorithm requires $2^N$ subtraction operations resulting in time complexity $O(2^N)$. This is demonstrated by the plot of the benchmarks shown in the plot in Figure 5. For set sizes $N < 8$, the CPU time is faster for the Classical algorithm. As the size of the set increases the advantage of INBL algorithm becomes more pronounced. The Classical algorithm required $O(2^N)$ hardware (memory) complexity, while the INBL algorithm required only O(400*$M$)=O($M$) value.

In conclusion, both the hardware and time complexity of the Classical algorithm were growing exponentially with $O(2^N)$, while the hardware complexity of INBL scaled polynomially by O($M$) and the time complexity was constant, O(1). Note: if the system of odd numbers would have been random/arbitrary instead of all the numbers then, paradoxically, the hardware complexity of both the Classical and INBL system would have been $O(2^N)$, where $2^N$ is the number of odd numbers.

Note, as we have mentioned at the beginning of this subsection, after carrying out the project described above, an additional solution, see Figure 6, by a classical Turing machine was proposed by (WD) in which each number in the set is represented by storing the higher-order bits in a bit set, omitting the least significant bit (LSB), which would be shared as a single flag for the entire set. To convert all numbers from odd to even (or vice versa), the shared LSB (of bit value 1) would simply be flipped (to zero), which updates the parity of the entire set with O(1) time complexity. Against this classical special-purpose algorithm, the INBL solution described above does not have superiority.

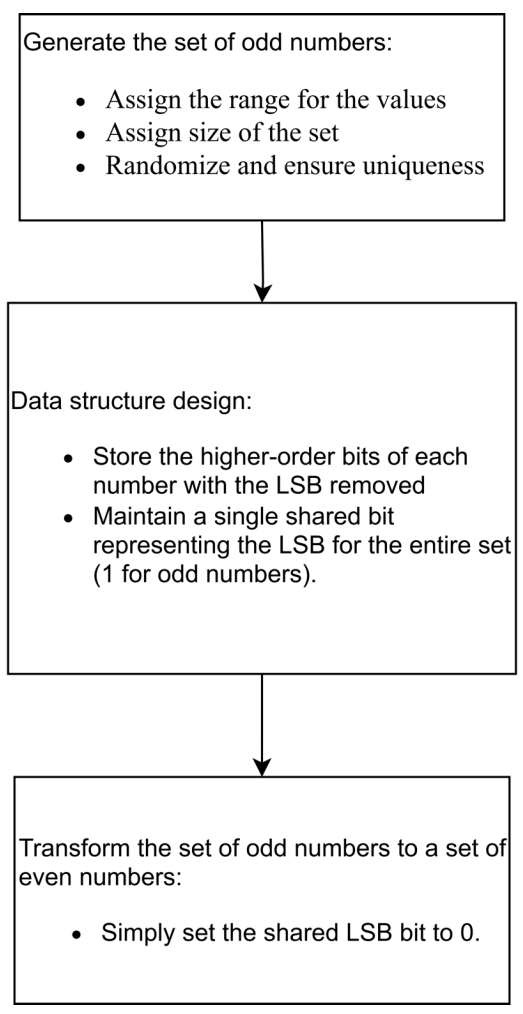

Fig. 6: WD's classical solution.

*4.2 Finding and removing a given number from a hat (unindexed database)*

We consider the problem [12] of removing a specific number $a \in S$ from an unsorted set $S = \{x_0, x_1, \dots, x_n\}$ of $n$ unique random numbers. The classical approach relies on a linear search algorithm to find the number, then the algorithm deletes the number at that index, resizing the underlying set and shifting all subsequent elements one position to the left to fill the gap. This process is sequential and it has linear time complexity versus the set size, which will increase exponentially with exponentially larger sets.

The INBL approach, see Figure 7, offers a fundamentally more efficient solution [12] both in terms of memory and time complexity.

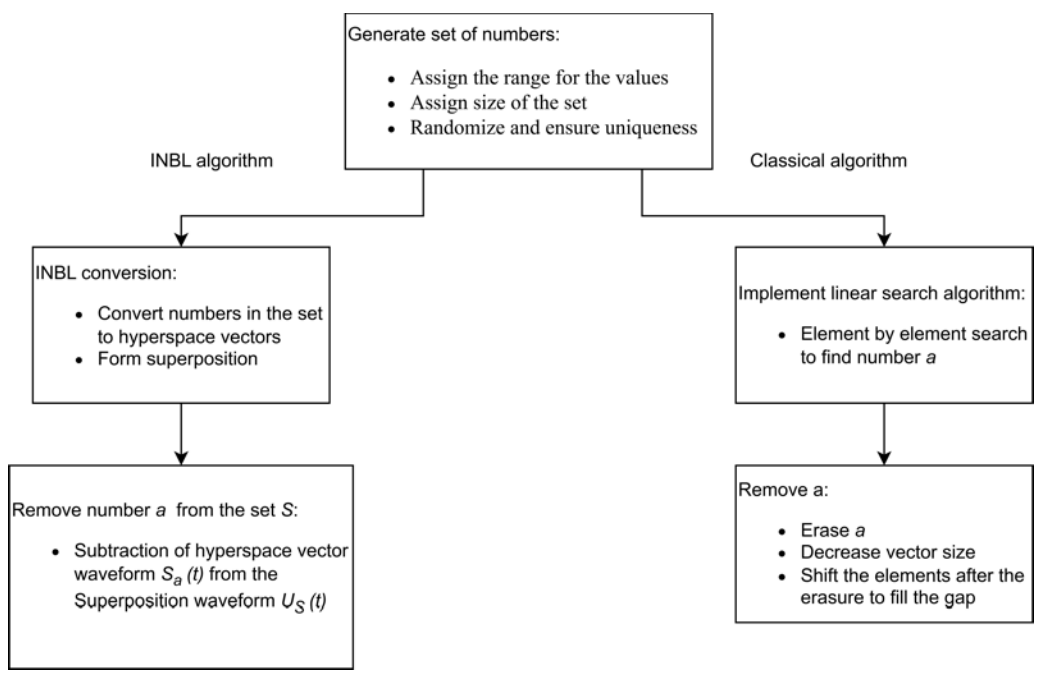

Figure 7. Finding and removing a given number from a heap (unsorted set, Hat). See the text for details.

Just as before using Equation 1, we generate the hyperspace vectors for each of the numbers in the set, and form the superposition:

$$U_S(t) = \sum_{k=0}^{2^N-1} S_k(t) \tag{9}$$

The hyperspace vector associated with number $a$ that will be removed is:

$$S_a(t) = \prod_{j=1}^{N} G_j^k(t) \tag{10}$$

Removing hyperspace vector $S_a$ from the superposition:

$$U_S^*(t) = U_S(t) - S_a(t) \tag{11}$$

In this simulation, we use RTWs with a duration of 100 clock cycles [20]. As a result, removing a hyperspace vector from the superposition requires exactly 100 subtraction operations. Importantly, this number remains constant regardless of the size or arrangement of the set, ensuring that the INBL algorithm maintains constant time complexity for the removal process.

Because numbers in an INBL superposition are inherently unordered, removing a specific number does not require any search algorithm. The removal operation is performed directly by subtraction, with constant time complexity, and no need for resizing or reorganizing the set.

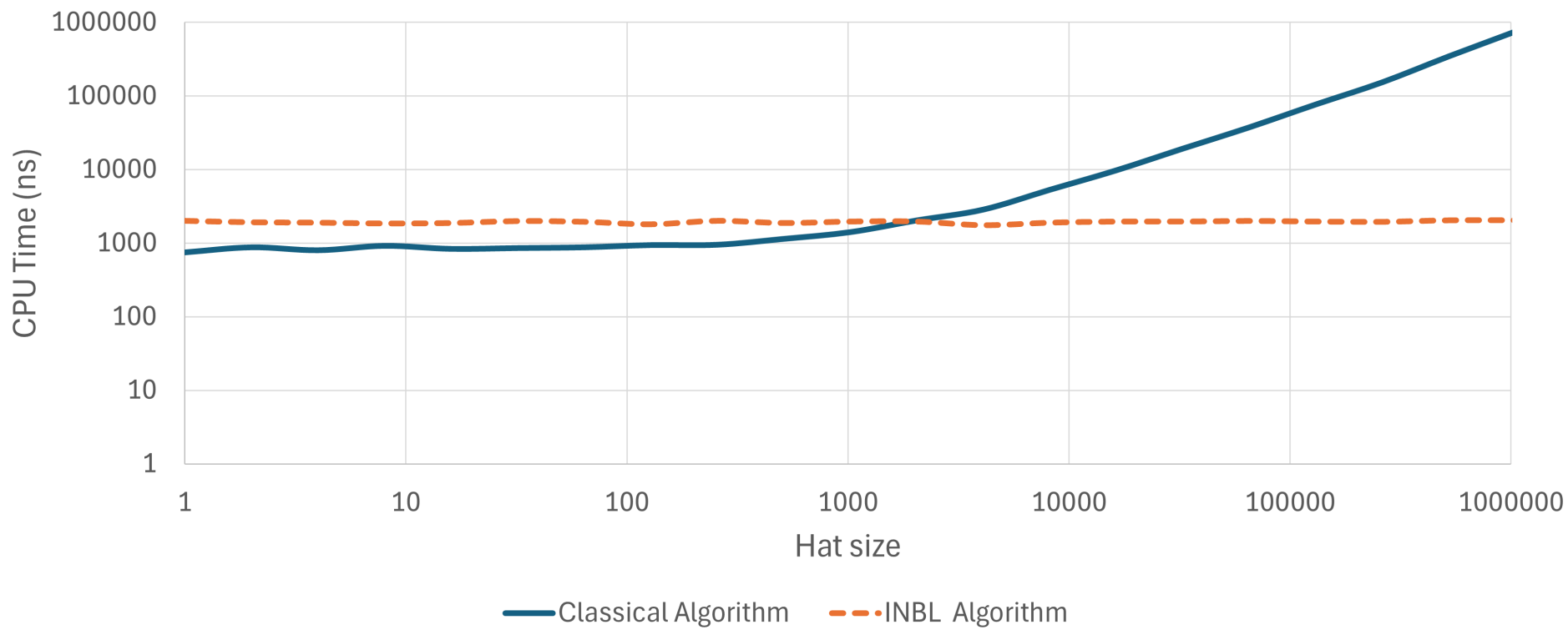

Figure 8. Finding and removing a given number from an exponential, unindexed set (Hat) [12].

## 5. Conclusions

The above described practical demonstration results quantify the exponential speedup by INBL for certain computational problems. Both the hardware and time complexity of the Classical algorithm were growing exponentially, $O(2^N)$, while the hardware complexity of INBL scaled only polynomially, $O(M)$, by the noise-bit resolution, and the time complexity was constant, O(1).

The observed exponential speed increase is in line with the theoretical advantages of INBL as a potential alternative to quantum computing. The INBL's ability to tackle such tasks with a low hardware and time complexity opens up possibilities for efficient solutions in scenarios where quantum computing may not be feasible.

## References

[1] L.B. Kish, ""Quantum supremacy" revisited: Low-complexity, deterministic solutions of the original Deutsch-Jozsa problem in classical physical systems", R. Soc. Open Sci. 10, 2023, 221327, https://doi.org/10.1098/rsos.221327
[2] S. Chen, "Quantum Advantage Showdowns Have No Clear Winners", Wired, July 11, 2022, https://www.wired.com/story/quantum-advantage-showdowns-have-no-clear-winners/ .
[3] R. Brierley "Not so fast", Nature Phys. 17, 2021, 1073. https://www.nature.com/articles/s41567-021-01388-9
[4] E. Tang, "Quantum Principal Component Analysis Only Achieves an Exponential Speedup Because of Its State Preparation Assumptions", Phys. Rev. Lett. 127, 2021, 060503.
[5] E. Tang "A Quantum-Inspired Classical Algorithm for Recommendation Systems", Proc. of the 51st Annual ACM SIGACT Symposium on Theory of Computing, STOC 2019, 2019, 217–228. https://doi.org/10.1145/3313276.3316310
[6] M. Sparkes, "Google's claim of quantum supremacy has been completely smashed", New Scientist, July 3, 2024, https://www.newscientist.com/article/2437886-googles-claim-of-quantum-supremacy-has-been-completely-smashed/ .

[7] L.B. Kish, "Noise-based logic: Binary, multi-valued, or fuzzy, with optional superposition of logic states", Physics Letters A 373 (2009) 911-918.
[8] L.B. Kish, S. Khatri, S. Sethuraman, "Noise-based logic hyperspace with the superposition of 2^N states in a single wire", Physics Letters A 373 (2009) 1928-1934.
[9] L.B. Kish, S. Khatri, S.M. Bezrukov, F. Peper, Z. Gingl, T. Horvath, "Noise-based deterministic logic and computing: a brief survey", International Journal of Unconventional Computing. 7 (2011) 101-113; https://doi.org/10.48550/arXiv.1007.5282
[10] L.B. Kish, W.C. Daugherity, "Noise-Based Logic Gates by Operations on the Reference System", Fluct. Noise Lett. 17 (2018) 1850033; DOI:10.1142/S0219477518500335
[11] Mohammad B. Khreishah, W.C. Daugherity, L.B. Kish, "XOR and XNOR Gates in Instantaneous Noise-Based Logic", Fluct. Noise Lett. 22 (2023) 2350041; DOI:10.1142/S0219477523500414
[12] B. Zhang, L.B. Kish, C. Granqvist, "Drawing From Hats by Noise-Based Logic", International Journal of Parallel, Emergent and Distribution System*s* (2015) http://arxiv.org/abs/1511.03552
[13] H. Wen, L.B. Kish, A. Klappenecker, "Complex Noise-Bits and Large-Scale Instantaneous Parallel Operations with Low Complexity", Fluct. Noise Lett. 12 (2013) 1350002.
[14] H. Wen, L.B. Kish, "Noise based logic: why noise? A comparative study of the necessity of randomness out of orthogonality", Fluct. Noise Lett. 11 (2012) 1250021.
[15] F. Peper, L.B. Kish, "Instantaneous, non-squeezed, noise-based logic", Fluct. Noise Lett. 10 (2011) 231-237. Open access: http://www.worldscinet.com/fnl/10/1002/open-access/S0219477511000521.pdf
[16] L.B. Kish, S. Khatri, F. Peper, "Instantaneous noise-based logic", Fluct. Noise Lett. 9 (2010) 323-330.
[17] Z. Gingl, S. Khatri, L.B. Kish, "Towards brain-inspired computing", Fluct. Noise Lett. 9 (2010) 403-412.
[18] S.M. Bezrukov, L.B. Kish, "Deterministic multivalued logic scheme for information processing and routing in the brain", Physics Letters A 373 (2009) 2338-2342.
[19] L.B. Kish, W.C. Daugherity, "Entanglement, and Unsorted Database Search in Noise-Based Logic", *Applied Sciences* 9 (2019) 3029; https://www.mdpi.com/2076-3417/9/15/3029/htm.
[20] L.B. Kish, S. Khatri, T. Horvath, "Computation using Noise-based Logic: Efficient String Verification over a Slow Communication Channel", *European Journal of Physics* B 79 (2011 January) 85-90; http://arxiv.org/abs/1005.1560
[21] L.B. Kish, C.G. Granqvist, S.M. Bezrukov, T. Horvath, "Brain: Biological noise-based logic", The 4th International Conference on Cognitive Neurodynamics, June 2013, Sigtuna, Sweden. Proceedings: Advances in Cognitive Neurodynamics (IV), H. Liljenström (ed.) Springer, (2014) pp. 319-322. DOI:10.1007/978-94-017-9548-7_45 . http://link.springer.com/chapter/10.1007%2F978-94-017-9548-7_45